# A universal bioluminescence tomography system for pre-clinical image-guided radiotherapy research

Zhishen Tong[1,†], Zijian Deng[1,†], Xiangkun Xu[1,†], Ciara Newman[1], Xun Jia[2], Yuncheng Zhong[2], Merle Reinhart[3], Paul Tsouchlos[3], Tim Devling[3], Hamid Dehghani[4], Iulian Iordachita[5], Debabrata Saha[6], James Kim[7], John W. Wong[2], and Ken Kang-Hsin Wang[1,*]

[1]Biomedical Imaging and Radiation Technology Laboratory (BIRTLab), Department of Radiation Oncology, University of Texas Southwestern Medical Center, Dallas, Texas, USA

[2]Department of Radiation Oncology and Molecular Radiation Sciences, Johns Hopkins University, Baltimore, Maryland, USA

[3]Xstrahl Inc., Suwanee, Georgia, USA

[4]School of Computer Science, University of Birmingham, Birmingham, B15 2TT, UK

[5]Laboratory for Computational Sensing and Robotics, Johns Hopkins University, Baltimore, Maryland, USA

[6]Department of Radiation Oncology, University of Texas Southwestern Medical Center, Dallas, Texas, USA

[7]Department of Internal Medicine, University of Texas Southwestern Medical Center, Dallas, Texas, USA

[†] The authors contributed equally to this work.

[*] Corresponding author:

Ken Kang-Hsin Wang, PhD

E-mail: kang-hsin.wang@utsouthwestern.edu

## Abstract

CBCT-guided small animal irradiators encounter challenges in localizing soft-tissue targets due to low imaging contrast. Bioluminescence tomography (BLT) offers a promising solution, but they have largely remained in laboratorial development, limiting accessibility for researchers. In this work, we develop a universal, commercial-graded BLT-guided system (MuriGlo) designed to seamlessly integrate with commercial irradiators and empower researchers for translational studies. We demonstrate its capabilities in supporting in vitro and in vivo studies. The MuriGlo comprises detachable mouse bed, thermostatic control, mirrors, filters, and CCD, enabling multi-projection and multi-spectral imaging. We evaluate that the thermostatic control effectively sustains animal temperature at 37°C throughout imaging and quantify that the system can detect as few as 61 GL261-AkaLuc cells in vitro. To illustrate how the MuriGlo can be utilized for *in vivo* image-guided research, we present 3 strategies, BLT-guided 5-arc, 2-field box, and BLI-guided single-beam, ranging from complicated high-conformal to simplest high-throughput plans. The high conformal BLT-guided 5-arc plan fully covers the gross tumor volume (GTV) at the prescribed dose with minimal normal tissue exposure (3.9%), while the simplified, high-throughput BLT-guided 2-field box achieves 100% GTV coverage but results in higher normal tissue exposure (13.1%). Moreover, we demonstrate that the localization accuracy of MuriGlo for both widely-used SARRP and SmART irradiators is within 1 mm, and the tumor coverage reaches over 97% with 0.75mm margin. The universal BLT-guided system offers seamless integration with commercial irradiators, achieving comparable localization accuracy, expected to supporting high-precision radiation research.

## Introduction

Pre-clinical radiotherapy(RT) research using animal models is an essential step to translate concepts from *in vitro* studies to clinical applications[1]. Small animal irradiators guided by computed tomography(CT) or cone-beam CT(CBCT), allowing precise radiation delivery, have transformed pre-clinical RT research[2-7]. While CT/CBCT provides excellent guidance capability, it is less adept at localizing soft tissue targets growing in a low-image contrast environment.

Bioluminescence imaging(BLI) has been widely used to visualize biological events by detecting BL signals emitted from organisms labeled with luciferase[8,9]. Due to its high image contrast, BLI offers an attractive solution for soft tissue targeting. However, two-dimensional(2D) BLI is limited in precisely localizing internal BL sources because the light propagation within tissues is susceptible to tissue optical properties and irregular animal torsos[10]. To tackle this challenge, we and others developed bioluminescence tomography(BLT) to reconstruct the 3D distribution of internal BL targets from surface BLI[10-15]. With such capability, BLT has been used to guide irradiation in various orthotopic tumor models, such as brain[11-13] , breast[14], and prostate[16], and also to monitor treatment response using BL intensity and tumor volume information[14]. However, such advanced image-guided systems remain in laboratorial development, not readily accessible to investigators for routine radiation studies.

To significantly enhance the conduct of research in scientific community and to complement the CBCT-guided irradiators[4,5], we collaborated with our industrial partner and developed a commercial-grade BLI/BLT system, MuriGlo. The optical platform was designed in a standalone setting with an innovative mouse bed system that enables seamless integration with various commercial irradiators to support image-guidance without requiring modifications to existing irradiators. To increase the data information for tomographic reconstruction and alleviate the underdetermined nature of BLT, we utilize multi-projection and multi-spectral imaging.

In this work, we demonstrate the capabilities of MuriGlo in supporting both *in vitro* and *in vivo* studies and provide a systematic workflow and treatment planning strategies to illustrate how BLT can be effectively utilized for *in vivo* image-guided radiation research. We also demonstrate that MuriGlo with the innovative mouse bed system, can be applied to widely-used commercial irradiators, small animal radiation research platform(SARRP, Xstrahl Inc, GA)[4] and small animal radiation therapy system(SmART, Precision X-ray Inc, CT)[5], to localize *in vivo* target for image-guided irradiation.

Specifically, we introduce a thermostatic system implemented in the imaging chamber to maintain the ambient temperature and its effectiveness in keeping the temperature at desired level for *in vivo* BL studies.

We also quantify the system sensitivity in detecting the minimal number of cells *in vitro* using luciferase-labeled murine glioblastoma(GBM) cell lines, GL261-*Luc2* and -*AkaLuc*[17]. The *Luc2* is commonly-used firefly reporter and the *AkaLuc* is the recently developed high brightness reporter[17]. To illustrate how the MuriGlo can be utilized for *in vivo* image-guided research, we present 3 strategies, BLT-guided 5-arc, 2-field box, and BLI-guided single-beam, ranging from complicated high-conformal to simplest high-throughput plans. Furthermore, we demonstrate that MuriGlo can lead to high precision *in vivo* image-guidance using orthotopic GBM model for both SARRP and SmART. The high precision and compatibility of this optical-guided system is expected to significantly enhance radiation research, offering functional targeting beyond anatomical imaging.

## Methods and Materials

### System configuration

MuriGlo consists of a rotational 3-mirror system, a fixed 45° mirror, filter wheel(Edmund Optics Inc., NJ), lens(50 mm, f/1.2, Nikkor, Nikon Inc., NY), and charge-coupled device(CCD) camera(iKon-M934, Andor Technology, Belfast, UK), enabling a compact imaging platform and multi-spectral BLT(Fig. 1). All mirrors are silver-coated with 98% reflectivity(H. L. Clausing Inc, IL). The BL signal emitted from imaged animal passes through the 3-mirror system, fixed mirror, filter, and is captured by the CCD. The 3-mirror can rotate 360° around the imaged object for multi-projection imaging. Four 20nm full-width at half-maximum band-pass filters(Chroma Technology Corp., VT) at 590, 610, 630 and 650nm were inserted into the filter wheel for multi-spectral imaging. Four LEDs(WS2812B 5050 RGB, Adafruit Industries LLC, NY) were mounted to the 3-mirror system to provide white-light illumination for imaging animals and fiducial markers on the mouse bed.

To minimize variation of BL cell spectrum caused by ambient temperature[18], an in-house thermostatic control was designed and installed in MuriGlo to maintain constant temperature for animal imaging. See Supplementary Sec.1 for thermostatic control description. To accurately measure the ambient temperature surrounding the imaged animal, a thermocouple was positioned on the rear surface of the 3$^{rd}$ mirror of the 3-mirror system(Supplementary Sec. 1 Fig. S1). Such design allows the thermocouple near the imaged mouse, ~5cm away from the mouse bed's center, and prevents collision between the thermocouple and bed during the 3-mirror rotation. Furthermore, two cooling fans(3610SB-05W-B30-B00, Digi-Key Electronics, MN) were installed in instrument chamber to circulate air and support CCD operation under -80°C while keeping imaging chamber at higher temperature(~37°C) for animal imaging(Fig. 1).

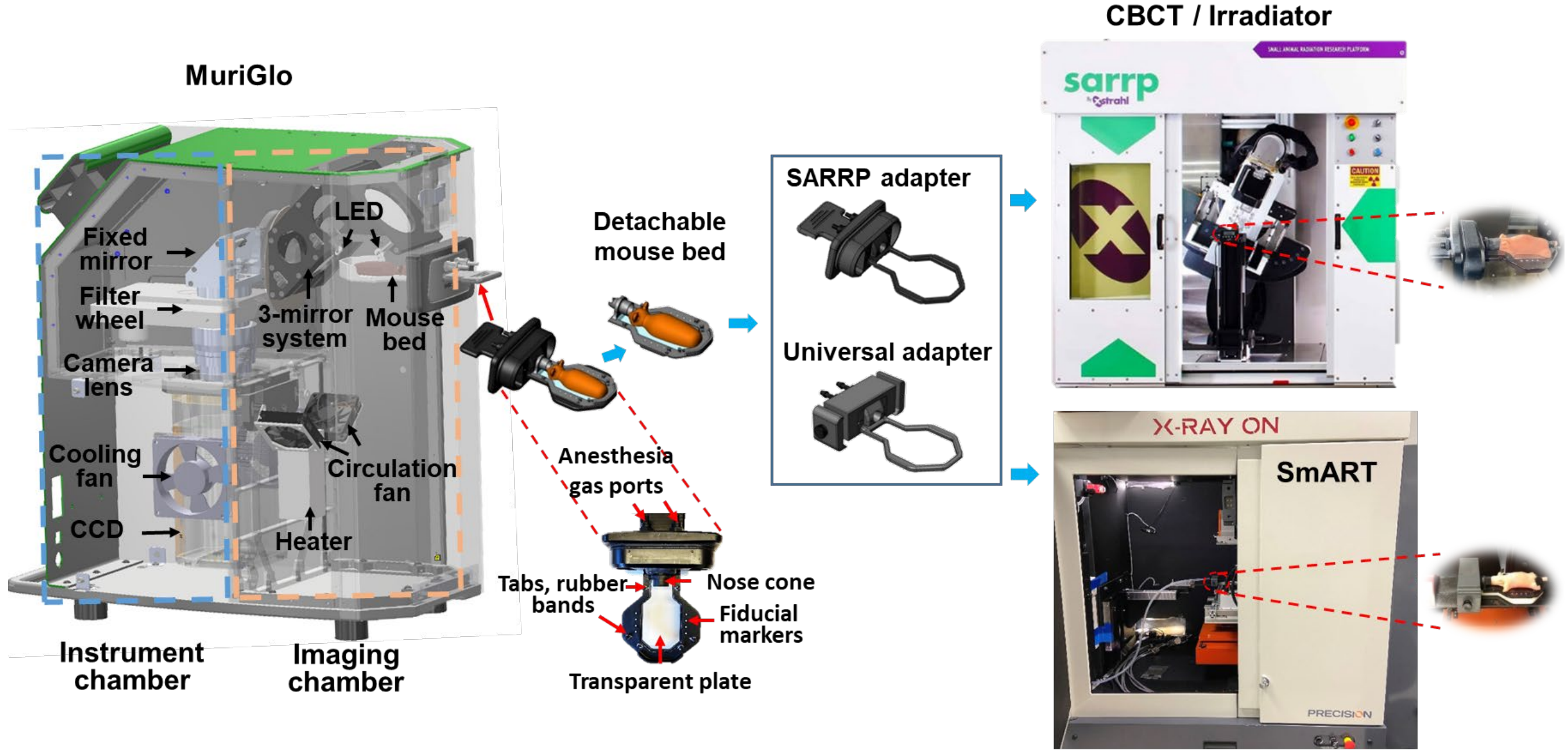

**Figure 1** illustrates the workflow of applying MuriGlo to guide irradiation for commercial irradiators. MuriGlo captures multi-spectral and multi-projection BL images from a mouse placed on a detachable bed with anesthesia and rubber bands for stabilization. Fiducial markers on the bed are employed for the CBCT and optical coordinate registration. The bed, with the anesthetized mouse, transfers from the bed adapter to a secondary adapter affixed to the irradiator for CBCT imaging and BLT-guided irradiation. Two types of secondary adapters were designed to accommodate commercial irradiators; the SARRP adapter is tailored for SARRP and the universal adapter can fit both SARRP and SmART.

For the BLT-guided process, tumor-bearing mouse is placed on the bed integrated with a bed adapter and inserted into MuriGlo for BL imaging acquisition. After imaging, the bed carrying anesthetized mice can be detached from the MuriGlo adapter and connected to secondary adapter affixed to SARRP or SmART irradiators for CBCT imaging, followed by BLT reconstruction for radiation guidance. Two types of secondary adapters are available; the $1^{st}$ kind can be directly attached to SARRP bed support, and the $2^{nd}$ kind named as universal adapter was designed to be placed on flat beds for both irradiators(Fig. 1). Throughout the image-guided session, the anesthetized animal is immobilized with rubber bands, and transport between MuriGlo and the irradiator is kept within 5m to minimize transportation process[10].

**Evaluating thermostatic control performance**

We conducted 3 tests to assess the performance of the thermostatic control in maintaining the ambient temperature of the imaging chamber and animal body temperature. In the first test, we checked whether the thermocouple(Sys-TC) on the 3-mirror system reflects the ambient temperature surrounding the animal. We placed a reference thermocouple(Ref-TC) at the bed's center to record the temperature on the bed(Supplementary Sec. 2 Fig. S2), and compared the reading of both TCs.

Since our BLT system was designed for single mouse imaging, to image multiple animals, after imaging one mouse, one would take the bed out from MuriGlo and replace next animal for imaging. For the second test, we aimed to investigate whether the chamber temperature would drop when removing the bed, and how quickly the bed can be warmed up to the desired temperature after it was inserted back to the system. To mimic the experiment workflow, we took out the bed attached with Ref-TC from MuriGlo, kept it outside for 10minutes, time for animal preparation, and then reinserted it into MuriGlo. Throughout this process, we monitored the time-resolved temperature changes for both Sys- and Ref-TC.

The third test assessed whether the thermostatic control could maintain the animal body temperature at ~37°C during imaging. We used rectal temperature as a measure of mouse body temperature. C57BL/6J albino mice(n=5, 6-8 weeks old, female; Jackson Laboratory, ME) were used. To measure the rectal temperature while anesthetized mouse on the bed, we placed a plastic-covered thermocouple into the mouse rectum with ~2cm depth. When the temperature reached ~36°C, the mouse was placed into MuriGlo, and we monitored body temperature to evaluate if thermostatic control maintained body temperature throughout imaging.

**System-specific cell spectrum**

For the multispectral BLT reconstruction, it is important to quantify the system spectral response and the emission spectrum of BL cells. For simplicity, we used the MuriGlo to measure the so-called system-specific spectrum, which combines the system and cell spectral response. For this study, we measured the spectrum of GL261-*Luc2* and -*Akaluc*(See Supplementary Sec. 3 for cell preparation). To prevent vaporization and heat loss from PBS, we used a 10-μm-thick polyvinyl chloride membrane, transparent to the spectrum of interest without attenuation, to cover the petri dish. Given the distinct spectral range of GL261-*Luc2* and -*Akaluc* cells, multispectral BLIs for GL261-*Luc2* at 590, 610, 630, and 650nm, and for GL261-*Akaluc* at 610, 630, 650 and 680nm were captured, respectively. To account for the dynamic *in vitro* BL signal variation throughout imaging, open-field images were taken before and after each spectral BLI, and an *in vitro* time-resolved signal curve was built. The intensity of multispectral BLIs taken at different time points was normalized to the time-resolved curve, and the normalized values as a function of wavelength resulted in the system-specific spectrum for the cell of interest. Additionally, to evaluate the impact of ambient temperature possibly on varying the spectrum, we compared two conditions: 23°C and 37°C, representing the MuriGlo with the thermostatic control deactivated and activated, respectively.

**Imaging sensitivity for *in vitro* study**

To assess the sensitivity of MuriGlo for *in vitro* imaging, we investigated the capability of its BLI to detect the minimal number of cells using GL261-*Luc2* and -*Akaluc* cells (See Supplementary Sec. 3 for cell preparation). The desired number of cells was achieved by cell dilution and subsequently placed in 96-well black plate. Experiments were repeated 3 times. Open-field BLIs without spectral filter at 10×10 binning and 60-second exposure time were used for the imaging acquisition. We used the signal-to-noise ratio(SNR)=5 as a threshold[19] to quantify the sensitivity of MuriGlo in detecting the minimal number of cells, defined as $\mathrm{SNR} = \langle S \rangle/\sigma$, where $\langle S \rangle$ is average counts of BL signal within the region of interest(ROI) and σ is the standard variation of background.

**Multispectral BLT reconstruction**

The forward model describing light propagation within tissue is given by diffusion approximation(DA) with the Robin-type boundary condition[20]. Specifically, the relationship between surface fluence rate $\boldsymbol{\varphi}_\lambda$ at wavelength $\lambda$ and internal source $\boldsymbol{S}$ can be expressed as a linear model:

$$\boldsymbol{G}_\lambda w_\lambda \boldsymbol{S} = \boldsymbol{\varphi}_\lambda, \tag{1}$$

where $\boldsymbol{G}_\lambda$ is the sensitivity matrix describing the changes of $\boldsymbol{\varphi}_\lambda$ related to the source $\boldsymbol{S}$, and $w_\lambda$ is the spectrum of the source $\boldsymbol{S}$. $\boldsymbol{G}_\lambda$ is calculated from the DA using the NIRFAST software based on finite element method(FEM) with the input of prior tissue optical properties[21]. Due to the non-contact imaging geometry, there is inevitable deviation between the actual surface fluence rate $\boldsymbol{\varphi}_\lambda$ and BLI measurement $\boldsymbol{b}_\lambda$. The spectral derivative(SD) method utilizing the ratio of neighboring wavelengths as the input for BLT reconstruction can minimize the deviation and eliminate complicated system modeling for non-contact imaging geometry[22]. The relationship between the $\boldsymbol{\varphi}_\lambda$ and $\boldsymbol{b}_\lambda$ is assumed as $\boldsymbol{\varphi}_\lambda = \boldsymbol{b}_\lambda \alpha$, where $\alpha$ is an unknown factor and assumed to be spectrally invariant. Additionally, considering the system spectral response for the acquisition $\boldsymbol{b}_\lambda$, the spectrum of BL source $w_\lambda$ is replaced by aforementioned system-specific spectrum $w_\lambda^s$. Accordingly, Eq.(1) becomes

$$\boldsymbol{G}_\lambda w_\lambda^s \boldsymbol{S} = \boldsymbol{b}_\lambda \alpha. \tag{2}$$

By applying the logarithmic form of Eq. (2) and the ratio of $\boldsymbol{b}_{\lambda_i}$ and $\boldsymbol{b}_{\lambda_{i+1}}$, Eq.(2) can be rewritten as

$$\left[\frac{\log \boldsymbol{b}_{\lambda_i}\alpha}{\boldsymbol{b}_{\lambda_i}\alpha} \boldsymbol{G}_{\lambda_i} w_{\lambda_i}^s - \frac{\log \boldsymbol{b}_{\lambda_{i+1}}\alpha}{\boldsymbol{b}_{\lambda_{i+1}}\alpha} \boldsymbol{G}_{\lambda_{i+1}} w_{\lambda_{i+1}}^s\right] \boldsymbol{S} = \log \frac{\boldsymbol{b}_{\lambda_i}}{\boldsymbol{b}_{\lambda_{i+1}}}. \tag{3}$$

The BL source distribution $\boldsymbol{S}$ in Eq.(3) can be iteratively solved by using a compressive sensing conjugated gradient algorithm[23].

***In vivo* BL GBM model and imaging process**

All animal procedures are in accordance with the institutional animal care and use committee at the University of Texas Southwestern Medical Center. We adopted the GBM model as a representative case to assess the capability of BLT system in localizing orthotopic tumors and to illustrate the planning strategies. To establish the GBM model, $1.2\times10^5$ GL261-*Luc2* cells were implanted into the left striatum of C57BL/6J albino mice(6-8 weeks old, female) at 3mm depth from the surgical opening. Two weeks after implantation, the GBM-bearing mice underwent multi-projections(-90°, 0° and 90°) and multispectral(610, 630 and 650 nm) BLI acquisition at 10×10 binning, initiated 10 minutes after D-Luciferin injection with concentration of 250μM/ml. The image taken at top of the bed was labeled as 0° projection imaging. Photo images at -60°, -30°, 0°, 30° and 60° projections were captured to retrieve fiducial marker positions for coordinate registration between BLIs and CBCT images[24]**Error! Reference source not found.**. Subsequently, the anesthetized mouse with the detachable bed was docked to the 2nd adapter pre-installed in irradiators for CBCT imaging. The CBCT image serves two purposes: 1) providing anatomical structure to generate a tetrahedral mesh used in the FEM-based BLT reconstruction, and 2) defining 3D coordinates in irradiators for treatment planning and irradiation.

The absorption coefficient $\mu_a$ 0.1610, 0.0820, and 0.0577$\text{mm}^{-1}$ and reduced scattering coefficient $\mu_s$' 1.56, 1.51, and 1.46 $\text{mm}^{-1}$ at 610, 630, and 650nm, respectively, and refractive index 1.4 were used for BLT reconstruction. A threshold of 0.5 of the maximum power density of BLT-reconstructed source was used to delineate the BLT-reconstructed gross target volume($GTV_{BLT}$)[12]. $T_2$-weighted fast spin echo sequence magnetic resonance imaging(MRI)(MRS 3017, 3T, MR Solutions Ltd. GU3 1LR, UK) was utilized to define the GTV of the GBM-bearing mice, serving as the ground truth to validate the accuracy of BLT localization. The MRI and CBCT image of mouse head were registered using 3D slicer[25].

**Treatment plans for *in vivo* GBM**

For the BLT-guided 5-arc plan, we first incorporated a uniform margin of 0.75mm accounting for the uncertainty associated with BLT target localization(e.g., target positioning and volume determination). This margin was added to the $GTV_{BLT}$ to form a planning target volume($PTV_{BLT}$). Subsequently, we designed a 5-arc BLT-guided plan using a motorized variable collimator(MVC) and MuriPlan treatment planning system(version 3.0.0; Xstrahl Inc.). The isocenter of each arc was set at the center of mass(CoM) of $GTV_{BLT}$ with the goal of 5Gy prescribed dose covering 95% of $PTV_{BLT}$. The 2-field plan was designed with 2 lateral beams and the isocenter was placed at the CoM of $GTV_{BLT}$. The collimation of each beam was opened 1mm beyond the edge of $GTV_{BLT}$ to account for the uncertainty of BLT target localization[12,24], resulting in 7×7 $mm^2$ as the collimator size for the *in vivo* case we present. Since tumor volumetric information would not be available when one uses surface BLI to guide irradiation, the 5×5$mm^2$ collimator, used in previous studies[11,12], was applied for the single beam to conservatively cover the two-week old GBM, <4 mm in diameter, based on our published tumor growth curve[11]. The central axis of the beam was aligned at the location where the maximum intensity of surface BLI is. The beam isocenter was placed at the inferior part of the skull to ensure sufficient coverage of GBM within the radiation field.

**Workflow for *in vivo* BLT-guided irradiation**

A summary workflow is provided in Fig. 2. For the data acquisition, multi-spectral and multi-projection BLIs are acquired, followed by multi-projection photo images for acquiring 2D body image and marker positions. The animal then undergoes CBCT imaging in irradiator. An in-house developed auto-contouring routine was adopted to generate body contour from the CBCT image(Supplementary Sec. 4). A geometrical calibration method published previously was used to register the 3D CBCT and 2D optical coordinates, based on the marker positions shownfrom both photo and CBCT image, enabling us map multi-projection BLIs onto the 3D mesh surface derived from the CBCT image[24]. The mapped surface BLIs serve as the input for BLT reconstruction. Given the BLT-reconstructed CoM, $GTV_{BLT}$ and $PTV_{BLT}$, we can devise image-guided treatment plans accordingly.

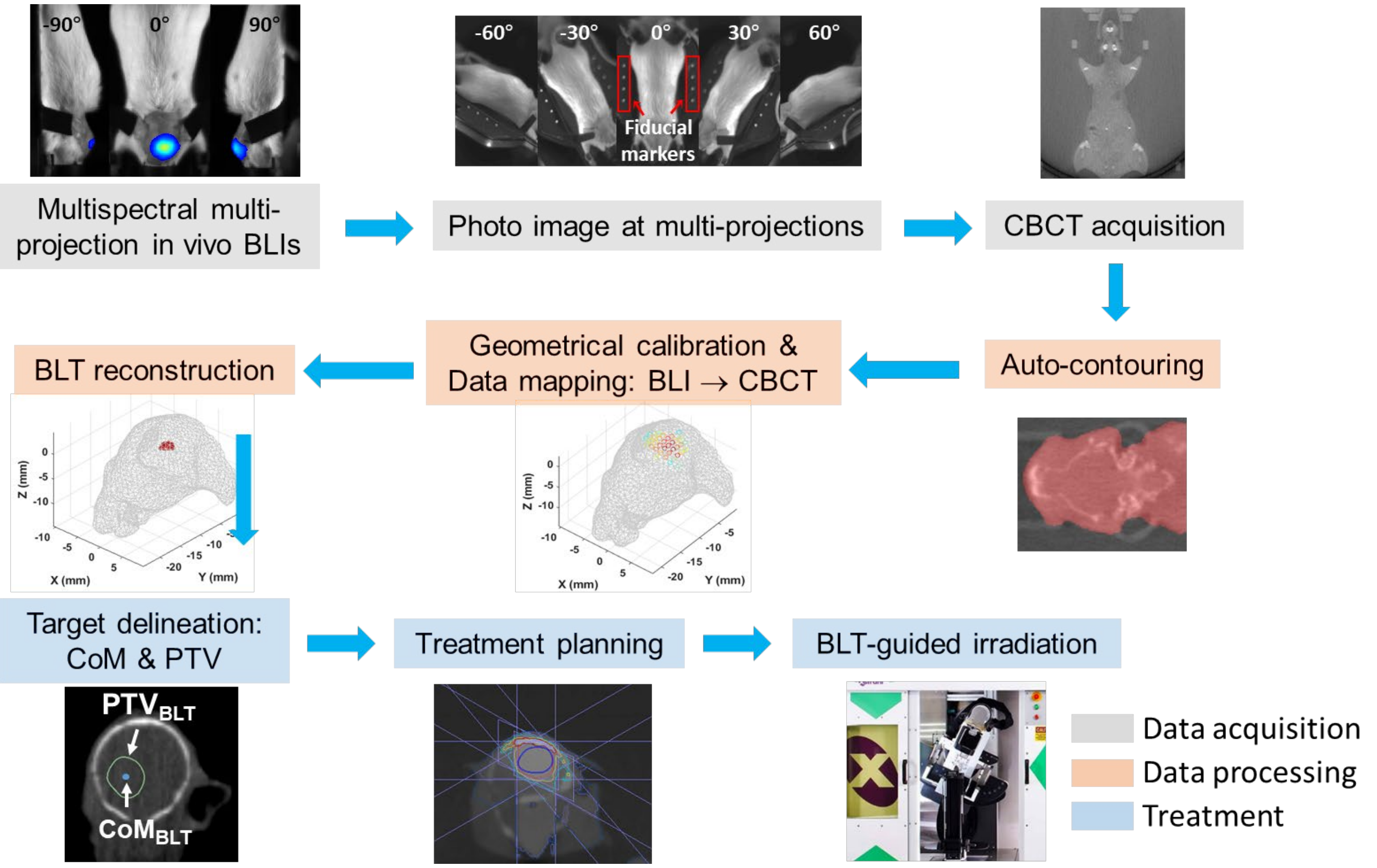

**Figure 2** shows the workflow of *in vivo* BLT-guided irradiation consisting of 1)data acquisition, 2)processing, and 3)irradiation. BL GBM model is used as an example shown.

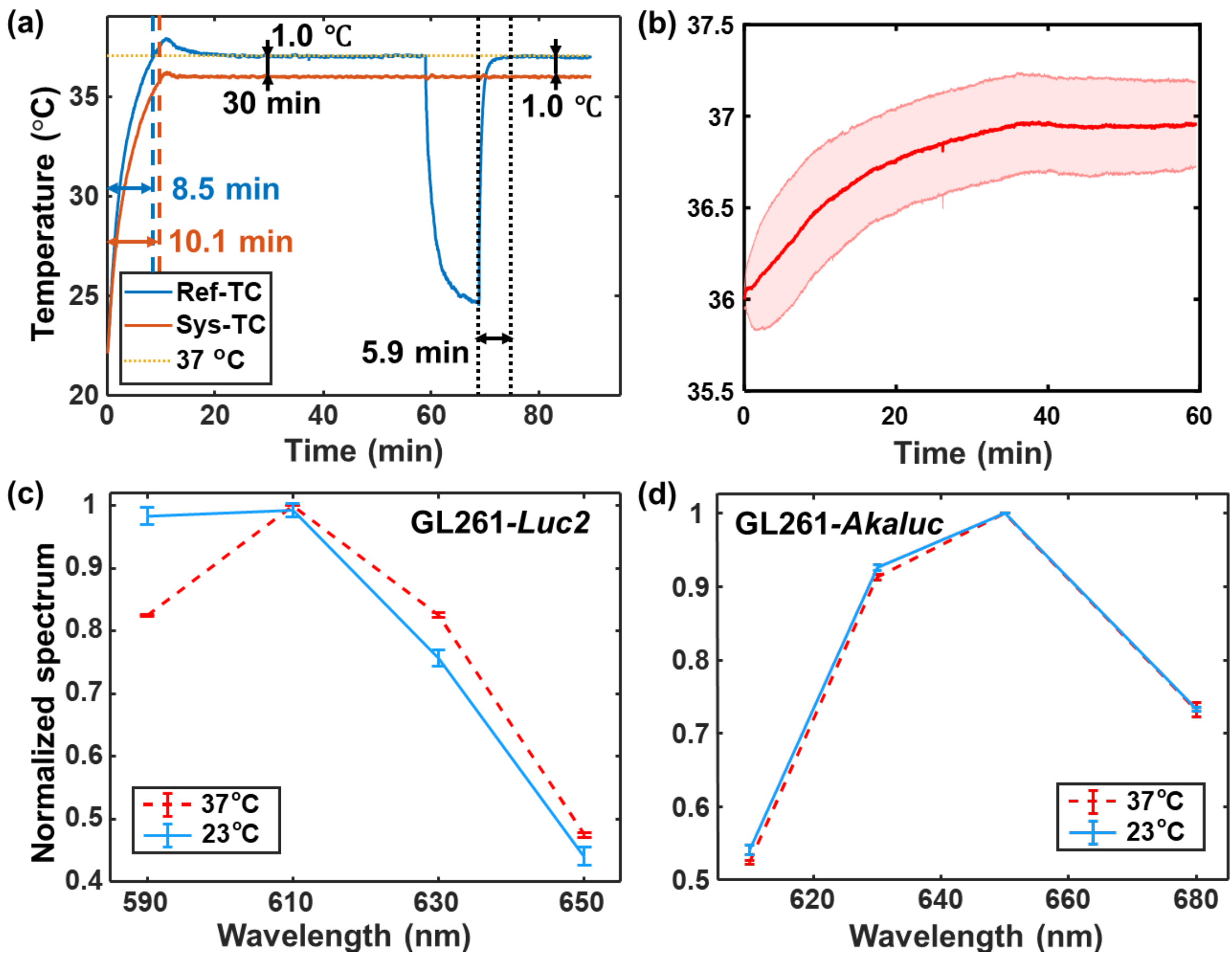

**Figure 3.** Assessment of thermostatic control and system-specific cell spectrum *in vitro* vs. ambient temperatures; (a) shows the reading from the Sys-TC and the Ref-TC. (b) shows measured rectal temperature vs. time(n=5); average: the red curve, std: pink area. (c) and (d) present the spectrum of GL261-*Luc2* and -*Akaluc* cells, respectively, at 37 and 23°C(n=6).

## Results

### Effectiveness of thermostatic control in maintaining chamber and animal temperature

As described previously, we used the Ref-TC placed on the bed to assess if the Sys-TC located on the 3-mirror accurately records the ambient temperature around the animal. Considering 37°C as the desired mouse body temperature and the distance between the Sys- and Ref-TC, we empirically set the PID temperature to 36°C. As shown in Fig. 3a, the Sys-TC could reach 36°C in 10.1minutes from room temperature after the heater was activated, while the Ref-TC arriving 37°C in 8.5minutes. After 30minutes, both TCs showed constant temperature at only 1°C difference and the Ref-TC shows 37°C around the bed. To reach the desired temperature at 37°C around animal and a thermal equilibrium state, a 30-minute warm-up period is suggested. Furthermore, to mimic the imaging workflow for multiple animals, we took out the bed attached with Ref-TC from MuriGlo at 60minutes, kept it outside for 10minutes, time for animal preparation, and then placed it back to the chamber. Despite the chamber opening introduced by the bed removal, the readings of Sys-TC remained stable throughout whole process and the Ref-TC reading indicates that the bed can be reheated to 37°C within 5.9minutes after reinsertion, confirming the effectiveness of the thermostatic control in maintain the chamber temperature.

We next assessed the capability of the thermostatic control in maintaining the animal body temperature. Figure 3b shows the averaged rectal temperature slightly increased from 36 to 37°C within 60minutes after system warmed up and animal placement, with the fluctuation within 0.7°C at 2 standard deviation(std).

**System-specific cell spectrum**

Figures 3c and d demonstrate the GL261-*Luc2* and -*Akaluc* cell lines have distinct spectral ranges. The system-specific spectrum of GL261-*Luc2* at 590, 610, 630 and 650nm at 37°C is 0.825±0.002, 1, 0.825±0.004, and 0.475±0.004(n=6), utilized as the input parameters for multispectral BLT reconstruction. It is worthwhile to note a red shift on the GL261-*Luc2* spectrum, when the ambient temperature increased from 23 to 37°C, while the spectrum of GL261-*Akaluc* was subject to minimum variation.

**Imaging sensitivity for *in vitro* bioluminescence study**

To assess the sensitivity of MuriGlo for *in vitro* imaging, we captured the open-field BLIs of GL261-*Luc2* and -*Akaluc* cells at given number of cells(Fig. 4a). We observed that the BL emitted efficacy of the GL261-*Akaluc* with Akalumine-HCL is about 21-fold higher than that of GL261-*Luc2* with D-Luciferin(Fig. 4a), aligned with previous findings[17] For both cell lines, the SNR increased linearly with the increase of cells. By using SNR=5 as a threshold, MuriGlo demonstrated the capability to detect as low as 1173 GL261-*Luc2* cells and 61 GL261-*Akaluc* cells, respectively(Fig. 4b).

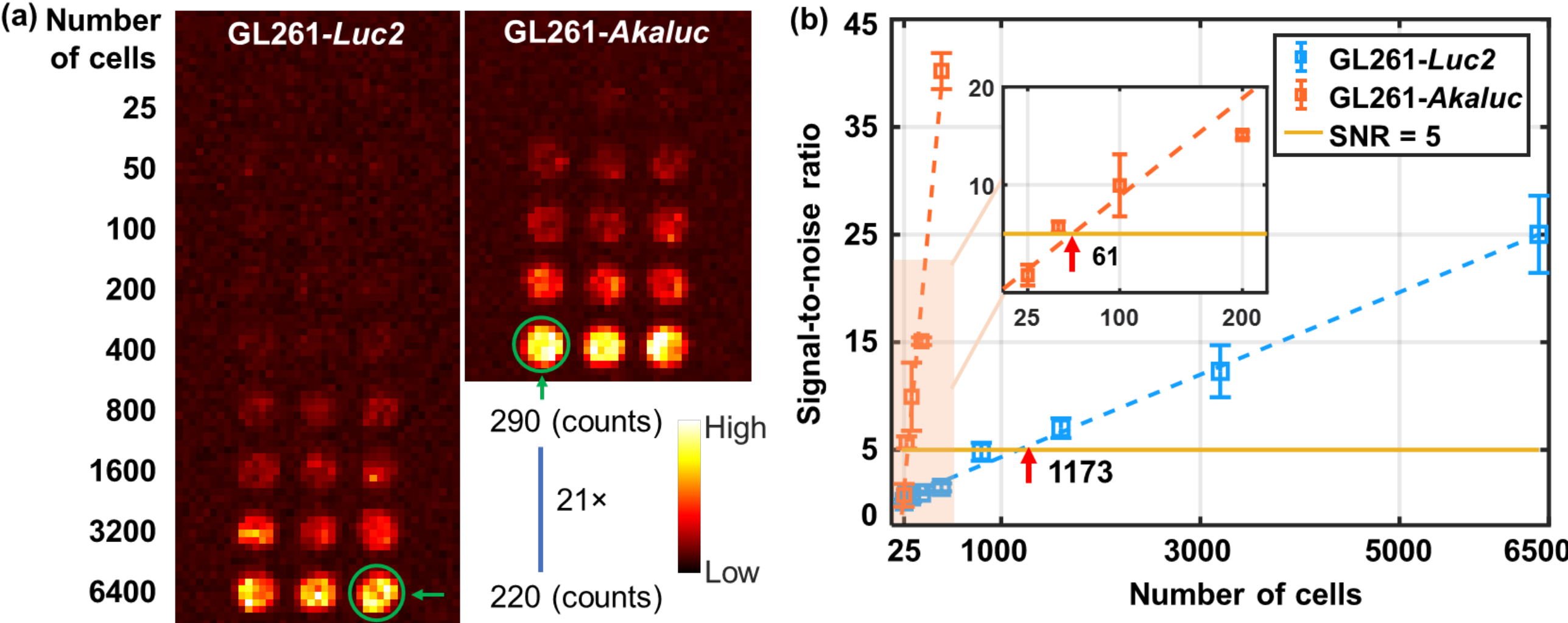

**Figure 4.** Imaging sensitivity of MuriGlo in detecting minimal numbers of GL261-*Luc2* and -*Akaluc* cells *in vitro*; (a) shows open-field BLIs vs. number of cells(n = 3). (b) shows the corresponding SNR vs. number of cells.

***In vivo* BLT and BLI-guided irradiation**

Figure 5a shows the BLI, originated from an *in vivo* GBM. Figure 5b is the corresponding BLI mapped onto the mesh surface of mouse head, used as the input data for BLT reconstruction. The $GTV_{BLT}$ of the BLT reconstructed tumor overlapped with the CBCT image is shown in Fig. 5c1–3. The tumor GTV delineated from the MRI(blue contour) was taken as the ground truth and registered to the CBCT image to compare with $GTV_{BLT}$. The average deviation between the CoM of $GTV_{BLT}$(black dots in Fig. 5c1–3, d1-3, and e3) and the geometry center of GTV is 0.61mm with std 0.37mm(n=4, Supplementary Sec.5). To account for the uncertainties of BLT target localization and volume delineation, we added a 0.75mm margin to $GTV_{BLT}$ to form the $PTV_{BLT}$, which can cover 100% of the GTV.

Figure 5c1-3 illustrates the BLT-guided 5-arc non-coplanar plan. With the prescribed dose of 5Gy covering 95% of the $PTV_{BLT}$, it demonstrates that the 5-arc conformal irradiation could fully cover the GTV while restraining normal tissue involvement. The 2-field box plan guided by the $GTV_{BLT}$ was proposed to demonstrate a high-throughput BLT-guided irradiation(Fig. 5d1-3). Compared to the 5-arc plan, the 2 field-plan could still fully cover the GTV at the prescribed dose with simplified radiation delivery, while larger normal tissue involvement is introduced. A BLI-guided single-field plan was also devised(Fig. 5e1-3) which represents the simplest and highest throughput radiation strategy among the

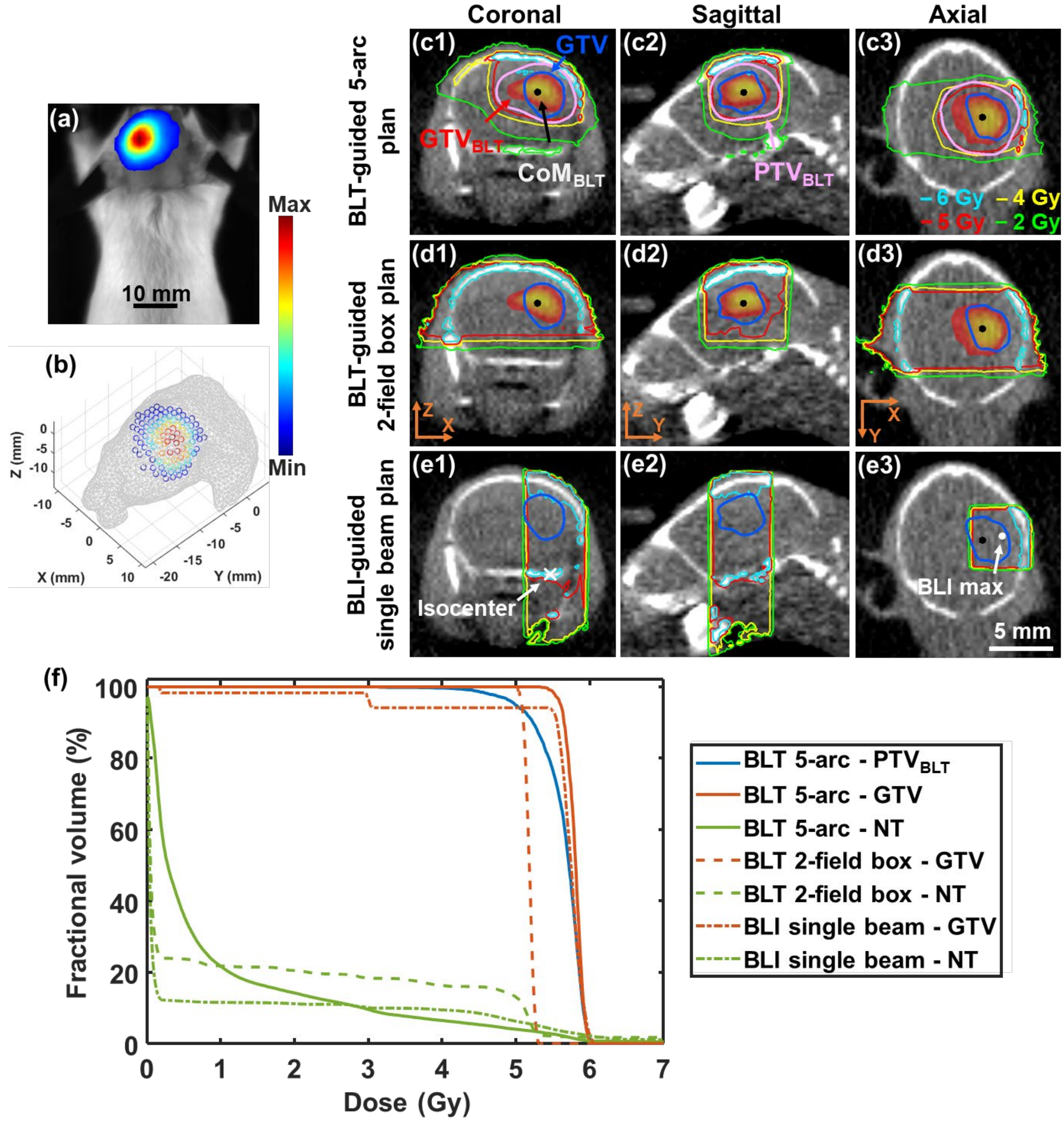

**Figure 5.** *In vivo* BLT reconstruction and BLT/BLI-guided radiation plan. (a) shows BLI(610nm; heat map) of a GBM-bearing mouse. (b) is the BLI mapped on mouse head mesh. (c1-3) is BLT-guided 5-arc plan with prescribed dose 5Gy covering the $PTV_{BLT}$(pink contour) generated from a BLT-reconstructed tumor, $GTV_{BLT}$(heat map), with MRI-resolved GTV resolved in blue contour and $GTV_{BLT}$ CoM as black point. (d1-3) is the 2-field box plan. (e1-3) is the single beam plan guided by BLI maximum(white dot, e3) and the isocenter was located at the skull(white cross, e1). (f) is the DVH of proposed plans for $PTV_{BLT}$, GTV, and normal tissue(NT).

three scenarios. However, because optical transport from an internal source is susceptible to irregular animal torso and tissue optical properties, the plan based on the BLI maximum position could not precisely guide the radiation beam, resulting in inferior tumor coverage.

Figure 5f shows the dose-volume histogram(DVH) of the three plans introduced in Fig. 5c-e. The blue solid curve indicates the 5Gy prescribed dose covering 95% of PTV$_{BLT}$ for the BLT-guided 5-arc plan. Both 5-arc and 2 field BLT-guided plans can achieve 100% GTV coverage. In contrast, the BLI-guided single-field plan can only cover 94% of the GTV at 5Gy, indicating inferior target coverage based on surface image. Furthermore, 3.9, 13.1 and 6.2% of normal tissue are involved within 5Gy shown in the BLT-guided 5-arc, 2-field box, and BLI-guided plan, respectively. The high conformality resulted from the 5-arc plans effectively constrains the volume involving>5Gy, leading to less radiation deposited to normal tissue.

**Integrating MuriGlo with the SmART irradiator**

To assess if one can apply MuriGlo to guide the SmART irradiator, we followed the procedures described in Figs. 1 and 2. The SmART CBCT was used for BLT reconstruction, determining the coordinate system for irradiation. The *in vivo* GBM was used to compare the BLT target localization between the MuriGlo-SmART and MuriGlo-SARRP integration. Figure 6a shows a representative case of the *in vivo* BLT reconstruction based on SmART CBCT. The GTV$_{BLT}$ is nicely overlapped with the GTV. The average deviation between the CoM of GTV$_{BLT}$ and the geometric center of GTV is 0.86mm with std 0.13mm(Fig. 6b) for the MuriGlo-SmART. By adding a 0.75mm uniform margin to GTV$_{BLT}$, the average tumor coverage by PTV$_{BLT}$ can reach 97.5%(Supplementary Sec.6). Figure 6b further demonstrates that the BLT target localization, CoM deviation, for both systems are<1mm, and the tumor can be covered by the PTV$_{BLT}$>97%. These results support that MuriGlo can be used as a universal optical system to guide commercially available irradiators.

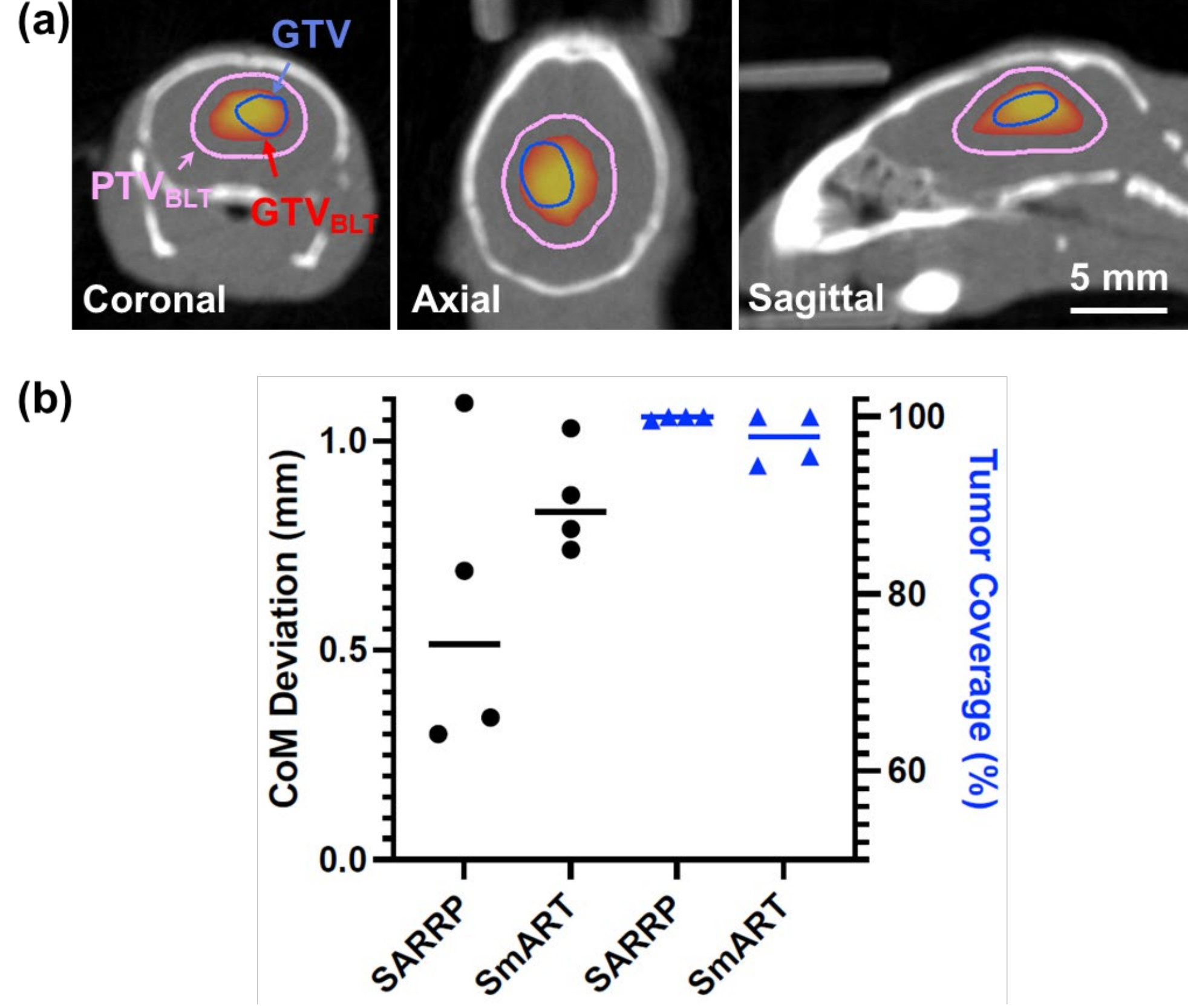

**Figure 6.** Comparison of *in vivo* BLT localization accuracy between MuriGlo-SmART and MuriGlo-SARRP. (a) is a representative case of applying MuriGlo to localize *in vivo* GBM based on SmART CBCT, showing overlap of SmART CBCT images GTV(blue contour), $GTV_{BLT}$ (BLT reconstruction, heat map), and $PTV_{BLT}$(pink contour). (b) is the CoM deviation and tumor coverage by $PTV_{BLT}$between the MuriGlo-SARRP and MuriGlo-SmART(n=4).

## Discussion

While CBCT provides excellent image-guidance capability, it is less adept at localizing soft-tissue targets growing in low-image contrast environments. Even with X-ray contrast agents, limitations remain due to their rapid clearance and applicability primarily in well-vascularized tumor models. Leveraging the high-contrast BLI, our group and others have developed BLT to reconstruct 3D target distribution, and applied it for irradiation guidance[10-16]. However, BLT-guided systems still largely remain in laboratorial development not readily accessible to investigators.

To meet these pressing needs, we collaborated with an industry partner to develop a commercial-grade BLI and BLT system, MuriGlo. The design of the system(Fig. 1) is a standalone unit featuring an innovative mouse bed system. This design not only ensures compatibility with current commercial irradiators for image-guided interventions without modifying existing equipment, but also facilitates BL imaging only applications independent of the irradiator. This approach significantly enhances both the optical system and the irradiator use. Commercial CBCT-guided systems like SARRP and SmART are transformative for pre-clinical RT research, with over 150 machines in use globally. MuriGlo's compact standalone configuration and high compatibility with existing irradiators provide dual advantages. First, it opens opportunities for institutions looking to advance their image-guided research capability. Second, it allows the substantial existing user base of SARRP and SmART to effortlessly incorporate the advanced optical-guided system, thus effectively enhancing their conduct of research. We compared the *in vivo* accuracy of the BLT localization for both SARRP and SmART(Fig. 6). The accuracy of target localization for MuriGlo-SARRP and MuriGlo-SmART configuration were within 1mm for the *in vivo* GBM model. With the small 0.75mm margin, we can reach similar tumor coverage>97% for both irradiators. These results confirm the optical-guided system can universally support both user groups in utilizing the emerging image-guided technique.

In this work, we also assessed our thermostatic system and illustrated why it is important for the BL application. The maintenance of normothermia in animal models is crucial for pre-clinical studies. The thermostatic control designed and implemented in the optical system can successfully sustain the animal body temperature at an average of 37°C with a minimal variation of 0.7°C(Figs. 3a and b). BL signal is the result of chemiluminescent reaction, and it has been observed that its emission efficiency(or brightness) can be influenced by ambient temperature[12,18] which impacts the imaging acquisition time and experiment throughput. In addition to the brightness, utilizing correct cell spectrum is equally important for the 3D tomographic application, because the multispectral BLT reconstruction relies on the spectrum as the input parameters. Thus, it is important to maintain a consistent spectrum between the *in vitro* measurement and *in vivo* BLT application. We observed the BL emission spectrum is both temperature and cell-line dependent(Fig. 3c vs d). Our results underscore the indispensability of precise temperature control for the BLI/BLT application to maintain normal physiological status of imaged animals and optimal BL emission and spectrum.

We further evaluated the sensitivity of the MuriGlo system for detecting minimal cell quantities by employing the widely-used BL reporter *Luc2* and the newly developed reporter *AkaLuc.* Our findings reveal that the MuriGlo system can detect as few as 61 GL261-AkaLuc cells. This observation suggests an approximate 21-fold increase in brightness compared to GL261-*Luc2*(Fig. 4a). Consistent with the findings of Iwano et al.[17] which highlighted the potential of *AkaLuc* and Akalumine-HCL system for single-cell BL imaging *in vivo* due to its superior brightness, our results establish a benchmark for the *in* *vitro* sensitivity of the MuriGlo. These insights are particularly valuable for researchers considering the adoption of MuriGlo for *in vitro* BL imaging applications. Furthermore, the development of brighter BL reporting systems, such as *AkaLuc*[17] and *NanoLuc*[26], enhances the potential of MuriGlo for effective deep tissue imaging.

Single-beam plan is commonly used in radiobiology studies[27-30] due to its throughput, but the normal tissue toxicity could confound the irradiation outcome. We show the BLT system could lead to highly conformal plan similar to clinical delivery(Fig. 5a1-3), in the cost of low throughput. Striking a balance between conformality and throughput is the simplified BLT-guided conformal plan i.e. 2-field box(Fig. 5d). By utilizing the 3D reconstructed tumor volume and considering the uncertainty of BLT target localization~1 mm[12,24](Fig. 6), we can design the collimator opening to effectively cover the tumor. Although this plan does not improve normal tissue sparing compared to the single beam approach(Fig. 5f), one could leverage the reconstructed volume to devise more conformal plan, like a 3 or 4-field box, to further reduce toxicity while at the cost of increased planning effort. This simplified BLT-guided planning approach could be considered when specific tumor margin data is unavailable, utilizing generic BLT localization uncertainties, such as 1mm in our case. Emerging techniques, i.e. sparse orthogonal collimator and inverse planning[31-1], are in development. The integration of these technologies and vendor support to streamline the planning procedures could enhance dose conformality, planning strategy, and throughput for pre-clinical RT research.

## Conclusion

The universal BLT system represents a significant innovation in pre-clinical RT research. The platform was demonstrated to be seamlessly integrated with commercially-available irradiators SARRP and SmART at comparable localization accuracy. The system can support both *in vitro* and *in vivo* studies. As we demonstrated the versatile planning strategies, investigators can utilize MuriGlo to realize high contrast optical imaging to support high precision radiation research.

## References

1. Marion DJ, Maina T. Of mice and humans: are they the same? —Implications in cancer translational research. *J Nucl Med* 2010; 51: 501-504.
2. Brown KH, Ghita M, Dubois LJ, et al. A scoping review of small animal image-guided radiotherapy research: advances, impact and future opportunities in translational radiobiology. *Clin Transl Radiat Oncol* 2022; 34:112-119.
3. Graves EE, Zhou H, Chatterjee R, et al. Design and evaluation of a variable aperture collimator for conformal radiotherapy of small animals using a microCT scanner. *Med Phys* 2007; 34: 4359-4367.
4. Wong J, Armour E, Kazanzides P, et al. High-resolution, small animal radiation research platform with x-ray tomographic guidance capabilities. *Int J Radiat Oncol Biol Phys* 2008; 71: 1591-1599.
5. Clarkson R, Lindsay PE, Ansell S, et al. Characterization of image quality and image-guidance performance of a preclinical microirradiator. *Med Phys* 2011; 38: 845-856.
6. Pidikiti R, Stojadinovic S, Speiser M, et al. Dosimetric characterization of an image-guided stereotactic small animal irradiator. *Phys Med Biol* 2011; 56: 2585-2599.
7. Verhaegen F, Granton P, Tryggestad E. Small animal radiotherapy research platforms. *Phys Med Biol* 2011; 56: R55-83.
8. Badr CE, Tannous B A. Bioluminescence imaging: progress and applications. *Trends Biotechnol* 2011; 29: 624-633.
9. Mezzanotte L, Root M, Karatas H, et al. In vivo molecular bioluminescence imaging: new tools and applications. *Trends Biotechnol* 2017; 35: 640-652.
10. Zhang B, Wang K H, Yu J, et al. Bioluminescence tomography–Guided radiation therapy for preclinical research. *Int J Radiat Oncol Biol Phys* 2016; 94: 1144-1153.
11. Deng Z, Xu X, Garzon-Muvdi T, et al. In vivo bioluminescence tomography center of mass-guided conformal irradiation. *Int J Radiat Oncol Biol Phys* 2020; 106: 612-620.
12. Xu X, Deng Z, Dehghani H, et al. Quantitative bioluminescence tomography-guided conformal irradiation for preclinical radiation research. *Int J Radiat Oncol Biol Phys* 2021; 111: 1310-1321.
13. Rezaeifar B, Wolfs C J A, Lieuwes N G, et al. A deep learning and Monte Carlo based framework for bioluminescence imaging center of mass-guided glioblastoma targeting. *Phys Med Biol* 2022; 67: 144003-144016.
14. Shi J, Udayakumar TS, Xu K, et al. Bioluminescence tomography guided small-animal radiation therapy and tumor response assessment. *Int J Radiat Oncol Biol Phys* 2018; 102: 848-857.
15. Chen J, Zhao N, Copello V, et al. Accurate and Early Metastases diagnosis in live animals with Multimodal X-ray and Optical Imaging. *Int J Radiat Oncol Biol Phys* 2023; 115: 511-517.
16. Shi J, Udayakumar TS, Wang Z, et al. Optical molecular imaging-guided radiation therapy part 1: Integrated x-ray and bioluminescence tomography. Med Phys. 2017; 44: 4786-4794.
17. Iwano S, Sugiyama M, Hama H, et al. Single-cell bioluminescence imaging of deep tissue in freely moving animals. Science 2018; 359: 935-939.
18. Rabha, MM, Sharma U, Barua AG. Light from a firefly at temperatures considerably higher and lower than normal. Sci Rep 2021;11: 12498.

19. Bushberg JT, John MB. The essential physics of medical imaging. *Lippincott Williams & Wilkins* 2011.
20. Wang LV, Wu HI. Radiative transfer equation and diffusion theory. *Biomedical Optics: Principles* and Imaging Wiley 2009; 93-118.
21. Dehghani H, Eames ME, Yalavarthy P K, et al. Near infrared optical tomography using NIRFAST: Algorithm for numerical model and image reconstruction. *Commun Numer Methods Eng* 2008; 25: 711-732.
22. Dehghani H, Guggenheim JA, Taylor SL, et al. Quantitative bioluminescence tomography using spectral derivative data. *Biomed Opt Express* 2018; 9: 4163-4174.
23. Basevi HR, Tichauer KM, Leblond F, et al. Compressive sensing based reconstruction in bioluminescence tomography improves image resolution and robustness to noise. *Biomed Opt Express* 2012; 3: 2131-2141.
24. Xu X, Deng Z, Sforza D, et al. Characterization of a commercial bioluminescence tomography-guided system for pre-clinical radiation research. *Med Phys* 2023; 50: 6433-6453.
25. Fedorov A, Beichel R, Kalpathy-Cramer J, et al. 3D Slicer as an image computing platform for the quantitative imaging network. *Magn Reson Imaging* 2012; 30: 1323–1341.
26. Su Y, Walker JR, Hall MP, Klein MA, Wu X, Encell LP, Casey KM, Liu LX, Hong G, Lin MZ, and Kirkland TA. An optimized bioluminescent substrate for non-invasive imaging in the brain. *Nat Chem Biol* 2023; 19: 731-739.
27. Zeng J, See AP, Phallen J, et al. Anti-PD-1 blockade and stereotactic radiation produce long-term survival in mice with intracranial gliomas. *Int J Radiat Oncol Biol Phy* 2013; 86: 343-349.
28. Zarghami N, Murrell DH, Jensen MD, et al. Half brain irradiation in a murine model of breast cancer brain metastasis: magnetic resonance imaging and histological assessments of dose-response. *Radiat Oncol* 2018; 13: 104.
29. Gunderson AJ, Yamazaki T, McCarty K, et al. Blockade of fibroblast activation protein in combination with radiation treatment in murine models of pancreatic adenocarcinoma. *PLoS One* 2019; 14: e0211117.
30. Andreou T, Williams J, Brownlie RJ, et al. Hematopoietic stem cell gene therapy targeting TGFb enhances the efficacy of irradiation therapy in a preclinical glioblastoma model. *J Immunother Cancer* 2021; 9: e001143.
31. Woods K, Nguyen D, Neph R, et al. A sparse orthogonal collimator for small animal intensity-modulated radiation therapy part I: Planning system development and commissioning. *Med Phys* 2019; 46: 5703–5713.
32. Woods K, Neph R, Nguyen D, et al. A sparse orthogonal collimator for small animal intensity-modulated radiation therapy. Part II: hardware development and commissioning. *Med Phys* 2019; 46: 5733–5747.
33. Balvert M, van Hoof SJ, Granton PV, et al. A framework for inverse planning of beam-on times for 3D small animal radiotherapy using interactive multi-objective optimization. Phys Med Biol 2015; 60: 5681–5698.

## Section 1 Description of thermocouple control system and setup

The thermocouple control system is composed of a type-T thermocouple (5TC-GG-T-30-36, Omega, CT), a proportion integration differentiation controller (PID, CN32Pt, Platinum Series, Omega, CT), a solid-state relay (SSRL240AC10, Omega, CT), a 24V transformer, a heater system including a heater (FCH-FGC15132R, Omega, Connecticut) and two circulation fans (CFM-6015V-254-362-20, Digi-Key Electronics, MN), a switch to control fans, two power distribution boards (PCB007; Evemodel, Jiangsu, China).

The type-T thermocouple is attached on the backside of the 3rd mirror in the 3-mirror system, which is the closest position to the imaged object, ~5 cm away from the center of mouse bed base (Figure S1a). To avoid the collision between the thermocouple and bed during the rotation of the 3-mirror system, we put the thermocouple 10 mm away from the symmetrical axis of the 3$^{rd}$ mirror (Figure S1b).

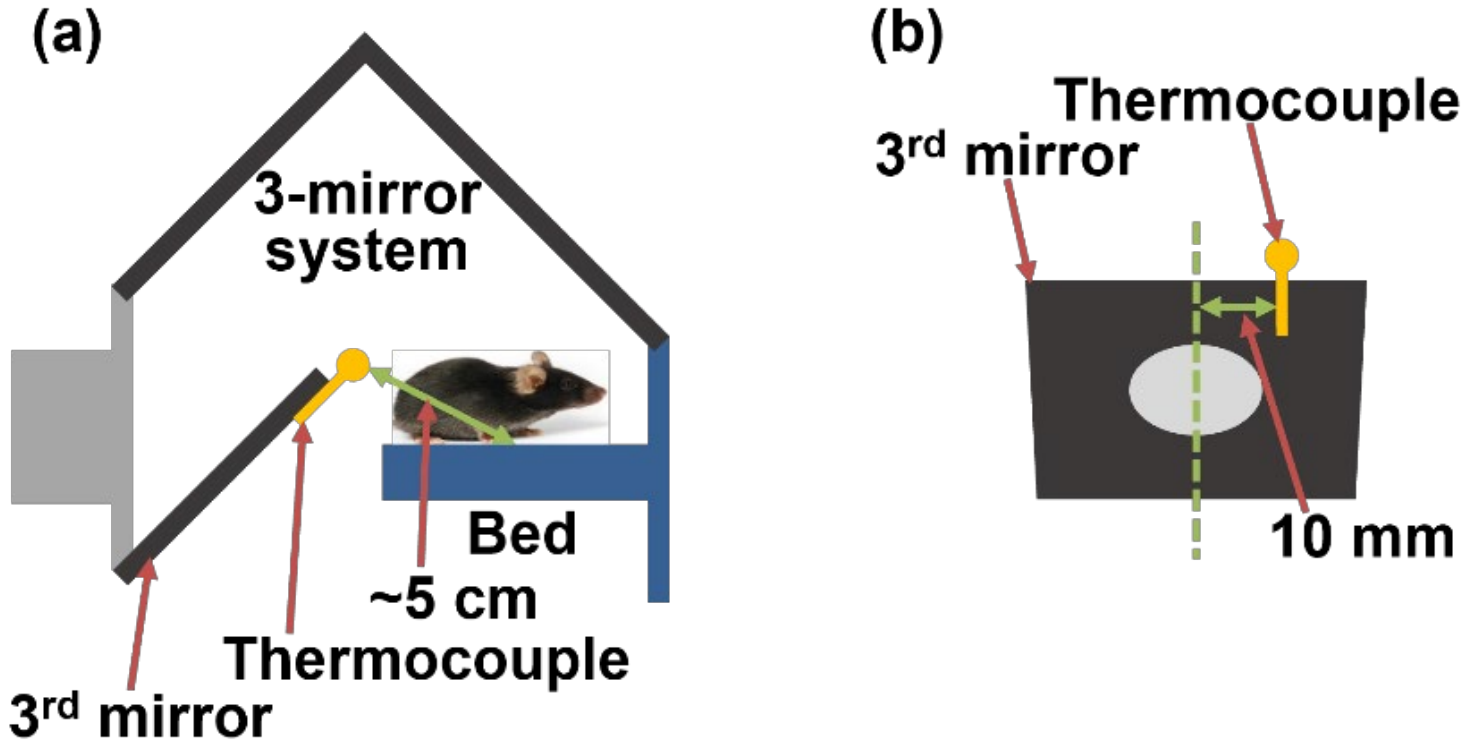

**Figure S1** (a) shows the position of thermocouple of thermocouple control system located at the backside of the 3$^{rd}$ mirror of the 3-mirror system, and (b) shows the front view of the 3$^{rd}$ mirror and the thermocouple was placed 10 mm away from its symmetry axis.

## Section 2 Reference thermocouple setup

To assess the deviation between system thermocouple measurement and temperature on the mouse bed, a reference thermocouple was placed on the bed center (Figure S2)

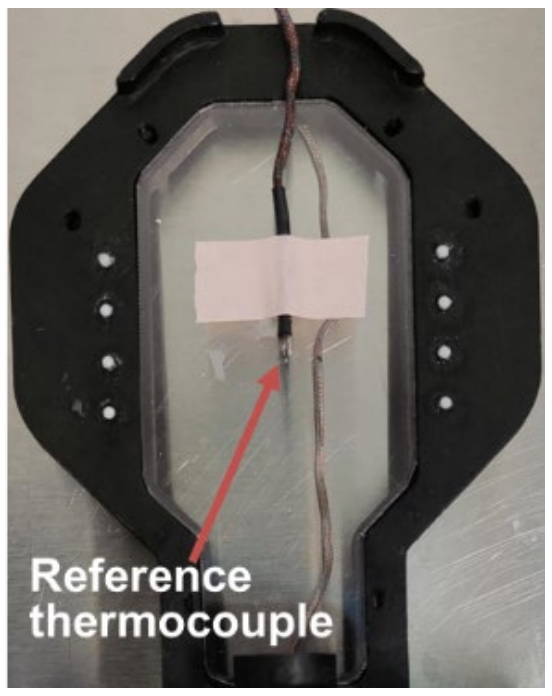

**Figure S2** The reference thermocouple is attached on the center of mouse bed to measure the temperature on the bed.

**Section 3 Cell preparation**

For system-specific spectrum measurements, both GL261-*Luc2* and -*Akaluc* cells were cultured in petri dishes, reaching a confluency level > 80%. GL261-*Luc2* cells were mixed with D-luciferin (PerkinElmer, MA) at a concentration of 0.75mg/mL in phosphate buffer solution (PBS) with pH=7.4, and GL261-*Akaluc* cells were mixed with Akalumine-HCL (Sigma-Aldrich, Darmstadt, Germany) at a concentration of 250µM/ml in PBS with same pH level for BL imaging.

To assess the sensitivity of MuriGlo in detecting the minimal number of cells *in vitro*, we prepared GL261-*Luc2* cells with a concentration 0.75 mg/ml of D-luciferin in 200µl Dulbecco's modified eagle medium (DMEM; Gibco™; Life Technologies Corp., NY), and GL261-*Akaluc* cells with a concentration of 250µM/ml of Akalumine-HCL in 200µl DMEM.

**Section 4 Auto-contouring**

To generate the animal body contour, we initially applied a Gaussian filter to the CBCT image to reduce scattering noise. We then defined a region of interest to exclude the mouse bed. To distinguish the mouse body from surrounding air, we employed a K-means method to generate an initial coarse contour and eliminated the nose cone from the coarse contour by using MATLAB subfunctions "imsegkmeans" and "imopen". Furthermore, we filled holes shown in the coarse contour, resulting from areas with similar image intensity to air, such as the trachea and lungs, using the MATLAB subfunction "imfill".

**Section 5 *In vivo* BLT reconstruction for MuriGlo-SARRP**

**5.1 Target localization**

Threshold 0.5 of the maximum reconstructed value was used to delineate BLT reconstructed gross tumor volume ($GTV_{BLT}$) and the center of mass (CoM) of $GTV_{BLT}$ was calculated. The deviation between the CoM of $GTV_{BLT}$, and the geometry center of $GTV_{MRI}$ delineated from MRI is shown in Table S1 for 4 mice. The CoM was calculated using the intensity of reconstructed BLT image, and the geometry center was calculated based on the MRI-delineated tumor geometry. Table S1 shows that the deviation between the CoM of $GTV_{BLT}$ and the geometry center of $GTV_{MRI}$ is 0.61 ± 0.37 mm, ranging from 0.30 mm to 1.09 mm for 4 mice.

**Table S1** Deviation between the CoM of $GTV_{BLT}$ and the geometry center of $GTV_{MRI}$

| Mouse label | M1 | M2 | M3 | M4 | Average ± Std |
|---|---|---|---|---|---|
| Deviation (mm) | 0.69 | 0.34 | 0.30 | 1.09 | 0.61 ± 0.37 |

## 5.2 Visualization of $GTV_{BLT}$ overlapped with $GTV_{MRI}$

In Figure S3, we present the overlap of $GTV_{MRI}$, and $GTV_{BLT}$ and $PTV_{BLT}$ resulted from MuriGlo integrated with SARRP. The $PTV_{BLT}$ with 0.75mm margin can cover 100% of the $GTV_{MRI}$.

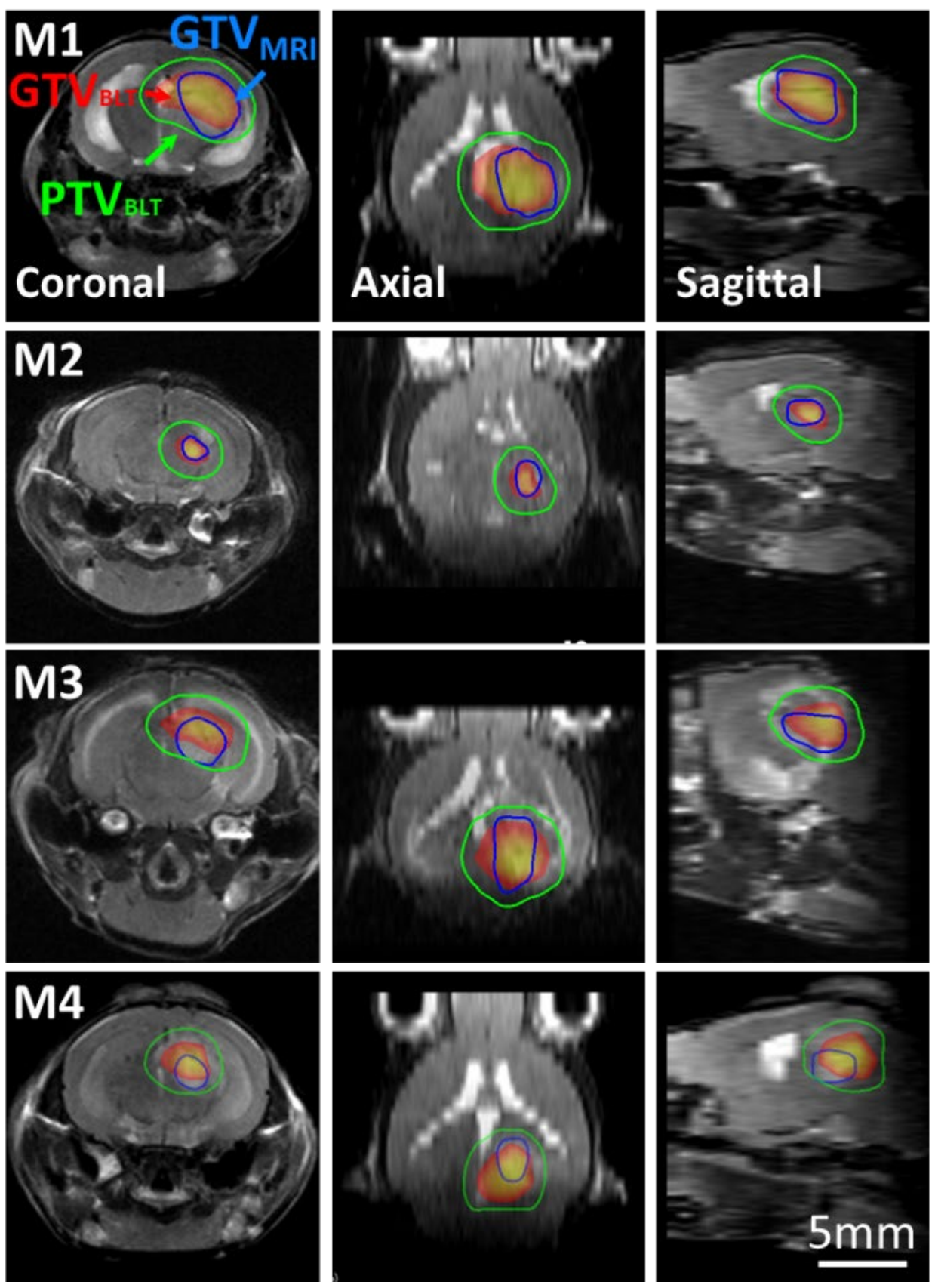

**Figure S3** The overlap of MRI-delineated GBM (gross target volume, $GTV_{MRI}$, blue contour), BLT reconstructed gross tumor volume ($GTV_{BLT}$, heat map) and MRI for 4 mice are shown in coronal, axial and sagittal views, respectively. A 0.75 mm margin was added to the $GTV_{BLT}$ to form $PTV_{BLT}$ (green contour).

## Section 6. *In vivo* BLT reconstruction for MuriGlo-SmART

## 6.1 Target localization

Table S2 shows the deviation between the CoM of $GTV_{BLT}$ and the geometry center of $GTV_{MRI}$ is $0.86 \pm 0.13$ mm, ranging from 0.74 mm to 1.03 mm for 4 mice.

**Table S2** Deviation between the CoM of $GTV_{BLT}$ and the geometry center of $GTV_{MRI}$

| Mouse label | M1 | M2 | M3 | M4 | Average ± Std |
|---|---|---|---|---|---|
| Deviation (mm) | 0.79 | 0.87 | 1.03 | 0.74 | 0.86 ± 0.13 |

### 6.2 Tumor and normal tissue coverage

The tumor coverage and normal tissue coverage are defined as $(PTV_{BLT} \cap GTV_{MRI})/GTV_{MRI}$ and $((PTV_{BLT}\text{-}PTV_{BLT}) \cap GTV_{MRI})/(V_{head}\text{-}GTV_{MRI})$, respectively, where the $V_{head}$ is the volume of mouse head. The tumor coverage and normal tissue coverage for different margins (0, 0.5, 0.75 and 1 mm) choices for 4 mice are shown in Figure S4. It shows that both tumor coverage and normal tissue coverage increase as the margin increases. Specifically, with 0.5 mm uniform margin added to $GTV_{BLT}$, compared to the 0 mm margin case, the average tumor and normal tissue coverages of $PTV_{BLT}$ are increased to 93.7% and 1.9%, respectively. With 0.75 mm uniform margin, the average tumor and normal tissue coverages are 97.5 % and 2.74%, respectively. With 1 mm uniform margin, the average tumor and normal tissue coverages are 99.2% and 3.79%, respectively. Considering the balance between tumor coverage and normal tissue coverage, the margin 0.75 mm was chosen for our study.

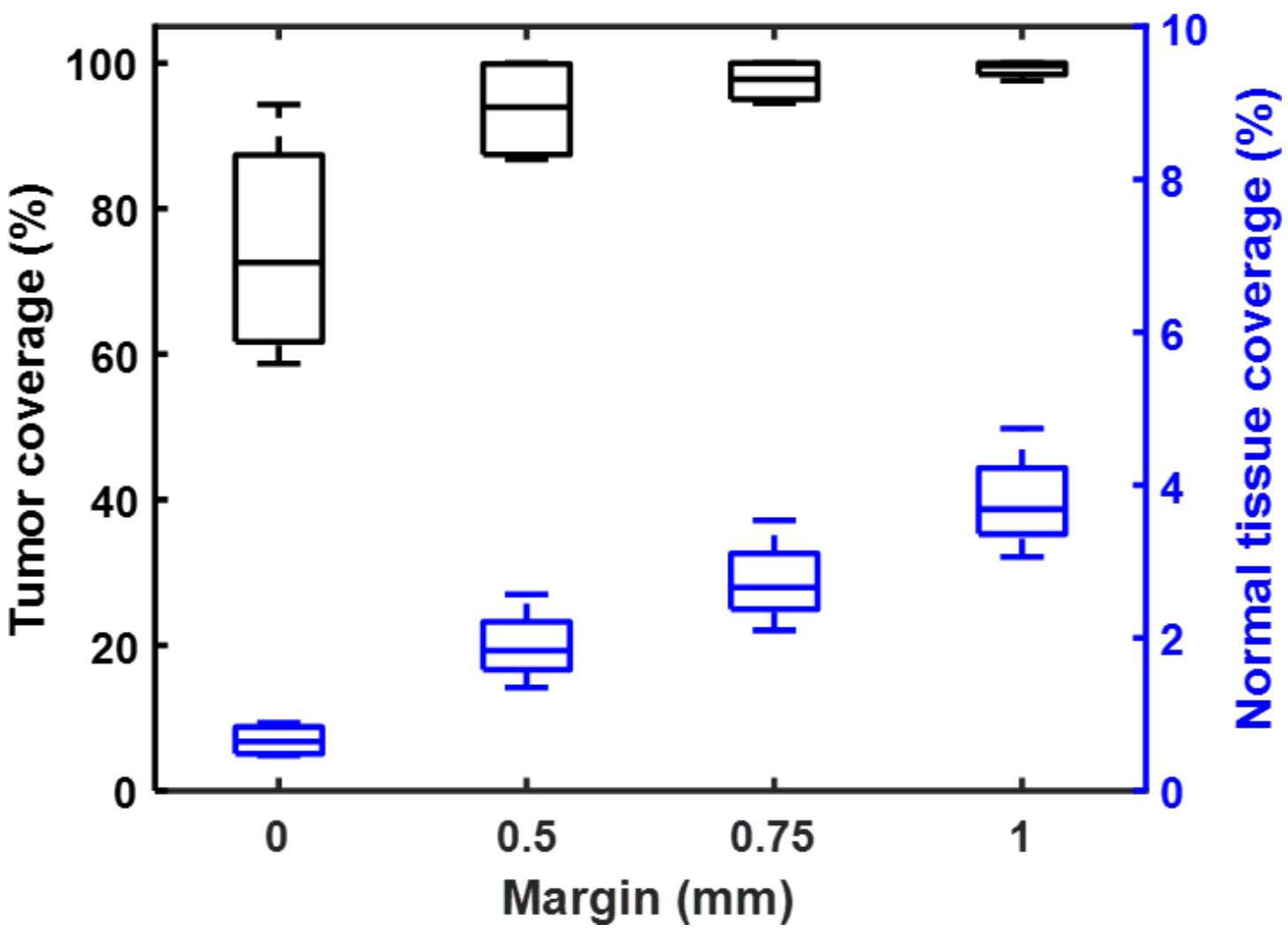

**Figure S4** Tumor coverage (left axis) and normal tissue coverage (right axis) versus margin expansion of 0, 0.5,0.75 and 1 mm for 2nd-week old GBM (n=4).

### 6.3 Visualization of $GTV_{BLT}$ overlapped with $GTV_{MRI}$

In Figure S5, we present the overlap of $GTV_{MRI}$, and the $GTV_{BLT}$ and $PTV_{BLT}$ resulted from MuriGlo integrated with SmART.

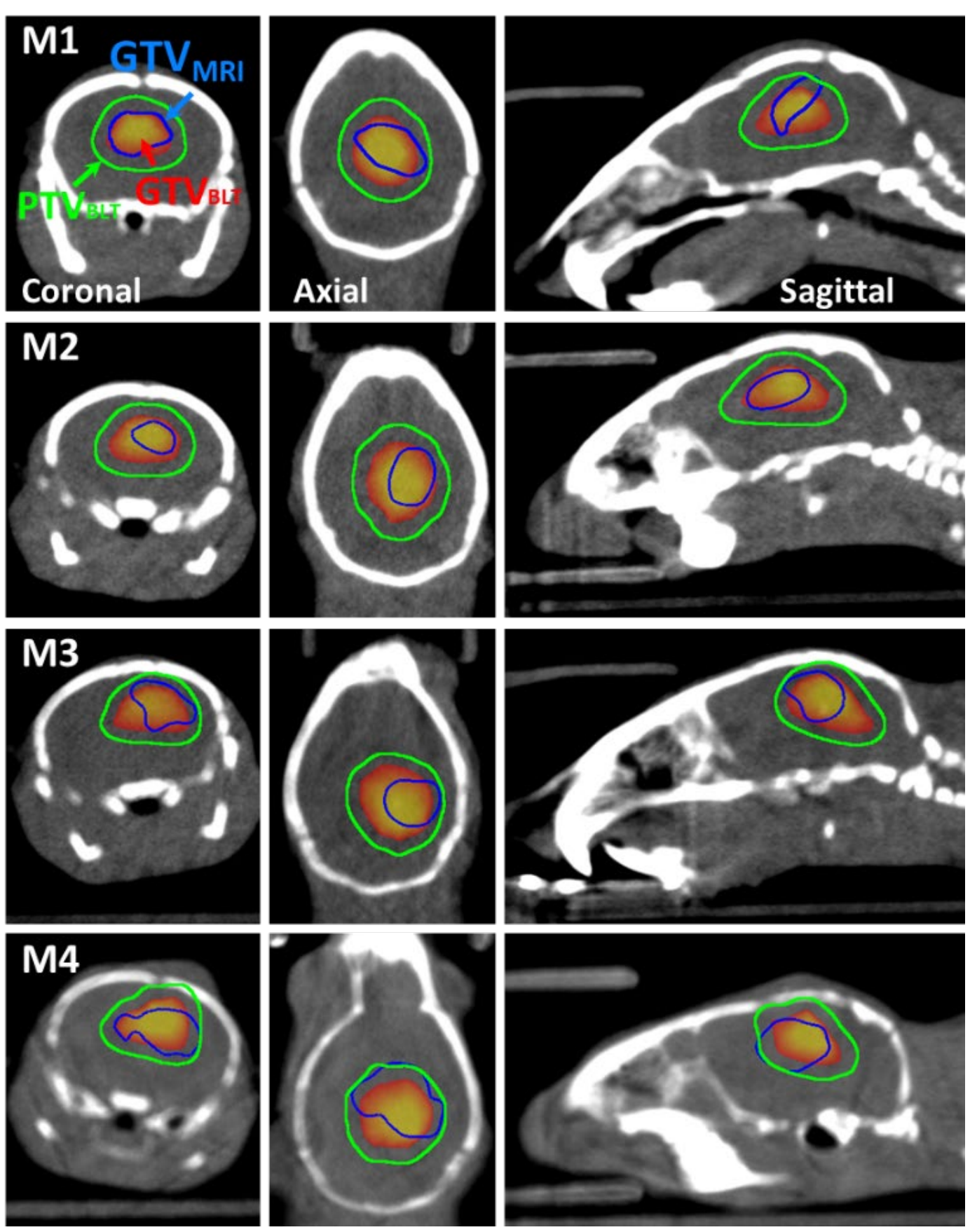

**Figure S5** The overlap of MRI-delineated GBM (gross target volume, $GTV_{MRI}$, blue contour), BLT reconstructed gross tumor volume ($GTV_{BLT}$, heat map) and CBCT for 4 mice are shown in coronal, axial and sagittal views, respectively. A 0.75 mm margin was added to the $GTV_{BLT}$ to form $PTV_{BLT}$ (green contour).